\documentclass[12pt]{article}
\usepackage{amsfonts}
\usepackage{latexsym}
\usepackage{dsfont}
\usepackage{amsmath}
\usepackage{amssymb}
\usepackage{amssymb}
\usepackage{amsthm}
\usepackage[nosort]{cite}
\usepackage[bulletsep]{collref}
\usepackage{graphicx}
\usepackage[font=small,labelfont=bf]{caption}
\usepackage{setspace}
\usepackage{alphabeta}
\usepackage[utf8]{inputenc}
\usepackage{geometry}
\usepackage[toc,page]{appendix}

\def\dd{\text{d}}
\def\tz{\tilde{z}}

\DeclareMathOperator\arcsinh{arcsinh}

\usepackage[hidelinks]{hyperref}
\hypersetup{
colorlinks=false,
citecolor= blue,
linkcolor= blue,
urlcolor= blue,
breaklinks=true
}

\def\appendix#1{\addtocounter{section}{1}\setcounter{equation}{0}
\renewcommand{\thesection}{\Alph{section}}
\section*{Appendix \thesection\protect\indent \parbox[t]{11.15cm}{#1}}
\addcontentsline{toc}{section}{Appendix \thesection\ \ \ #1}}

\allowdisplaybreaks

\numberwithin{equation}{section}

\begin{document}

%%%%%%%%%%%%% TITLE %%%%%%%%%%%%%%%%%%%%%%%%%

\begin{titlepage}
\begin{center}
%\today
\vspace*{-1.0cm}
\hfill {\footnotesize HU-EP-21/02,  DMUS-MP-21/01}

\vspace{2.0cm}

{\LARGE  {\fontfamily{lmodern}\selectfont \bf Light-Cone Gauge
in \\
\vspace{3mm}
Non-Relativistic AdS$_5\times$S$^5$ String Theory}} \\[.2cm]

\vskip 1.5cm
\textsc{Andrea Fontanella\footnote{\href{mailto:afontanella@perimeterinstitute.ca}{\texttt{afontanella@perimeterinstitute.ca}} \\
Current affiliation: Perimeter Institute for Theoretical Physics,
Waterloo, Ontario, N2L 2Y5, Canada. 
} \footnotesize
and \normalsize  Juan Miguel Nieto Garc\'ia\footnote{\href{mailto:juan.miguel.nieto.garcia@desy.de}{\texttt{juan.miguel.nieto.garcia@desy.de}} \\ Current affiliation: II. Institut für Theoretische Physik, Universität Hamburg,
Luruper Chaussee 149, 22761 Hamburg, Germany.}}
\\
\vskip 1.2cm

\begin{small}
\textit{$^1$Institut f\"ur Physik, Humboldt-Universit\"at zu Berlin, \\
IRIS Geb\"aude, Zum Gro{\beta}en Windkanal 2, 12489 Berlin, Germany}

\vspace{5mm}

\textit{$^2$Department of Mathematics,
University of Surrey \\
Guildford, GU2 7XH, UK. }
\end{small}

\end{center}

\vskip 0.7 cm
\begin{abstract}
\vskip0.5cm

\noindent We discuss the non-relativistic limit of string theory in AdS$_5\times$S$^5$ for different choices of embedding coordinates. We show that, if we consider Cartesian coordinates, the action of fluctuations around the twisted BMN-like string found in \href{https://arxiv.org/abs/2109.13240}{arXiv:2109.13240} in uniform light-cone gauge becomes the one of free fields at large string tension and large AdS$_5$ radius.

\end{abstract}

\end{titlepage}

\tableofcontents
\vspace{5mm}
\hrule

%%%%%%%%%%%%%%%%%%% BODY %%%%%%%%%%%%%%%%%%

 \setcounter{section}{0}
 \setcounter{footnote}{0}

\section*{Introduction}\addcontentsline{toc}{section}{Introduction}

Recently, we experienced in the community of theoretical high-energy physics a growing interest regarding the topic of non-relativistic string theory.  The first theory of non-relativistic strings was formulated in flat spacetime \cite{Gomis:2000bd,Danielsson:2000gi} and later it was also constructed in AdS$_5\times$S$^5$ \cite{Gomis:2005pg}, but it is only recently that we gained a better understanding of these type of theories. 
In \cite{Bergshoeff:2018yvt, Bergshoeff:2019pij} it was found that non-relativistic string theories (in flat or curved backgrounds) are theories secretly defined on a manifold which is \emph{String Newton-Cartan}.  This is quite different from the usual relativistic string theory, where the string is free to move on a Riemannian manifold. These articles also discuss T-duality and symmetries for non-relativistic string theories\footnote{In these articles, only closed strings have been considered.  For recent work on non-relativistic \emph{open} strings, see \cite{Gomis:2020fui, Gomis:2020izd}.}.  The approach used in \cite{Bergshoeff:2018yvt, Bergshoeff:2019pij} consists in expanding the relativistic vielbein at large parameter $\omega$, which plays the stringy analogue role of the speed of light, and extract the String Newton-Cartan geometry data as coefficients of the expansion. Before taking $\omega$ to infinity, this procedure requires us to couple the relativistic action to a closed critical B-field, in order to trade a divergent term in the metric for additional non-dynamical fields, which can be interpreted as Lagrange multipliers.   

Alternative approaches to the limit procedure have been proposed. One of them, called the \emph{null-reduction} approach, consists in obtaining a Polyakov-type action on a torsional Newton-Cartan geometry (of a particle type) from the reduction of a Riemannian background with a null isometry \cite{Harmark:2017rpg, Harmark:2018cdl,Harmark:2019upf}. 
Another approach, called the \emph{expansion} approach and proposed in \cite{Hartong:2021ekg, Hartong:2022dsx}, consists in expanding the relativistic action in a large parameter $\omega$ and study the equations of motion up to a certain order in $\omega$, without having to take a limit on $\omega$, and therefore without having to introduce a critical B-field. The idea behind the expansion approach is reminiscent of the Lie algebra expansion method applied to coset string sigma models \cite{Fontanella:2020eje}, where in the latter one it is required a concrete target space isometry algebra in order to apply the expansion on it.
In this article, we shall take the limit point of view.

Non-relativistic string theories have many differences with their relativistic counterparts. Here we highlight two of them. First, non-relativistic string theories cannot be consistently defined on a Lorentzian background. Despite that, when they are defined on a String Newton-Cartan background, but with a world-sheet that remains relativistic, these theories are free of Weyl anomalies \cite{Gomis:2019zyu,Gallegos:2019icg} and they can be regarded as a consistent string theories, independent of their parental relativistic ones, to be studied on their own.  
Second, these theories require having a winding mode to be consistent. As it was pointed out in flat space, in order to obtain a non-empty spectrum in a non-relativistic string theory, one needs to impose that the string  wraps around the longitudinal spatial direction of the background geometry. This happens also for non-relativistic AdS$_5\times$S$^5$ string theory, as it was found to be a consistency condition imposed by the Lagrange multipliers' equations of motion \cite{Fontanella:2021btt, Fontanella:2023men}.

In view of those characteristics, one could ask what could be the non-relativistic version of the AdS/CFT duality. So far, such duality has been largely studied in the relativistic case, where the most famous version is the one that relates the AdS$_5\times$S$^5$ relativistic string theory and $\mathcal{N}=4$ SYM theory in 4d \cite{Aharony:1999ti}. To begin the program of testing a non-relativistic version of the duality, in the string theory side one needs to compute the spectrum of the string excitations. This has been done with great success in relativistic AdS$_5\times$S$^5$ string theory, where integrability in the planar limit has played an important role in terms of making computations manageable \cite{Beisert:2010jr}.

Progress towards understanding classical integrability in the non-relativistic AdS$_5\times$S$^5$ string action has recently been made.\footnote{Another instance where integrability appears manifestly is by taking a further non-relativistic limit by sending the world-sheet speed of light to infinity. In this way, one ends up with a theory which has a non-relativistic world-sheet and target space geometry. When this is applied to AdS$_5\times$S$^5$ string theory, the double scaled theory is called  Spin Matrix theory, which in its simplest case is the Landau-Liftshitz theory \cite{Harmark:2017rpg, Harmark:2018cdl, Harmark:2019upf}, and it is known to be integrable.  A generalisation to other values of the charges has been studied in \cite{Harmark:2020vll}.} A coset description of the String Newton-Cartan AdS$_5 \times$S$^5$ manifold was found in \cite{Fontanella:2022fjd, Fontanella:2022pbm}, which made possible to formulate a non-relativistic version of the Metsaev-Tseytlin coset action. Thanks to this formalism, it was possible to find a Lax pair \cite{Fontanella:2022fjd}, as a first step towards proving classical integrability.   
The spectral curve associated with such Lax pair was studied in \cite{Fontanella:2022wfj} and the non-standard result obtained indicates that the usual notion of spectral curve should be generalised to take into account the non-diagonalisability of the monodromy matrix in the context of non semi-simple isometry algebras, which is the case when taking the non-relativistic limit.

The spirit of this article is to start applying ideas developed in relativistic AdS$_5\times$S$^5$ string theory \cite{Arutyunov:2009ga, Arutyunov:2005hd,Frolov:2006cc}, which are useful in order to test the AdS/CFT duality,  also to the non-relativistic theory.  In order to find the full non-relativistic spectrum of string excitations, one possible way is to find the S-matrix of the theory and then apply Bethe ansatz techniques. To determine the S-matrix, first we need to fix a gauge in such a way that we get rid of the redundant string degrees of freedom.  In the context of relativistic AdS$_5\times$S$^5$ string theory, the \emph{light-cone gauge} is particularly useful, since in this gauge one can implement all Virasoro constraints directly in the action, having no ghosts to deal with, and in the large string tension limit the action expands perturbatively around the BMN vacuum as a 2d free massive scalar fields at quadratic level. This is particularly convenient for quantisation and for computing the S-matrix governing scattering of string excitations \cite{Beisert:2010jr,Arutyunov:2009ga, Arutyunov:2005hd,Frolov:2006cc}. 

At the moment, the spectral problem of non-relativistic AdS$_5\times$S$^5$ string theory has been studied only in the \emph{static gauge}, see e.g. \cite{Sakaguchi:2007zsa,Sakaguchi:2007ba}. In the static gauge, the action of quadratic fluctuations describes 3 massive and 5 massless free scalar fields propagating in AdS$_2$.  Although the dynamics of these fields is ``free'', there is still an important complication due to the fact that they propagate in a curved 2d spacetime (i.e. AdS$_2$), and finding the S-matrix of scalar fields moving in AdS$_2$ is notoriously difficult,  see e.g.  \cite{Anninos:2019oka}. 
Moreover, a perturbative expansion in large AdS$_2$ radius would produce untamed terms at the asymptotic infinities, preventing us from defining asymptotic scattering states, as one would usually do in flat spacetime. This is something that usually does not happen in light-cone gauge, as typically the action expands about free scalar fields in 2d flat spacetime with tamed correction terms, such that, in general, asymptotic scattering states are well-defined.

With that goal in mind, in this article we plan to tackle the following two points:
\begin{itemize}
    \item  \emph{Non-relativistic limit in different coordinates.} 
We address the derivation of the non-relativistic AdS$_5\times$S$^5$ string theory starting from different choices of embedding coordinates. The idea is to start with the relativistic string action written in different set of coordinates, which are all equivalent. This raises the question of how to take a sensible non-relativistic limit in all these different coordinates. We show that the rescaling of the coordinates can be fixed by the \.In\"on\"u-Wigner contraction of the AdS$_5\times$S$^5$ isometry algebra, in particular via the choice of coset representative defining the specific set of coordinates considered. As we shall see, non-relativistic theories derived from different set of coordinates lead to different sets of String Newton-Cartan data. We comment whether these sets are equivalent by means of gauge transformations.

\item \emph{Semi-classical expansion.} 
We show that the Cartesian coordinates analogue of the non-relativistic theory proposed in \cite{Gomis:2005pg}, after fixing uniform light-cone gauge\footnote{We point out that the Hamiltonian formulation for actions based on String Newton-Cartan backgrounds has been studied in \cite{Kluson:2018grx}. Moreover, light-cone gauge has been fixed for a theory obtained from rescaling $t$ in AdS$_5$ and $\phi$ in S$^5$ \cite{Kluson:2017ufb}, which however does not single out an AdS$_2$ subspace inside AdS$_5$ as in \cite{Gomis:2005pg} and in our case.} and when the vacuum is the twisted periodic BMN-like string found in \cite{Fontanella:2021btt}, admits a perturbative expansion around free fields in large string tension and large AdS$_5$ radius.

\end{itemize}

This article is organised as follows.  In section \ref{sec:non_rel_general} we introduce the notion of non-relativistic string theory based on a general String Newton-Cartan geometry. The idea of the derivation may be regarded as quite general, although in some points our discussion will be closely referring to the case of AdS$_5\times$S$^5$.  In this section, we also discuss what happens when one takes different sets of coordinates for the initial relativistic theory, and we shall discuss that in general one lands on theories with different String Newton-Cartan data.
In section \ref{sec:lighcone} we fix uniform light-cone gauge for the bosonic sector of non-relativistic AdS$_5\times$S$^5$ string theory, and we show that the action expands in large string tension and large AdS$_5$ radius around the twisted periodic BMN-like vacuum as a free fields action in Mink$_2$. In section \ref{sec:conclusions} we will summarise our results and give future prospects. This article closes with two appendices. In Appendix \ref{Appx_NCdata} we define the algebra contraction, the sets of coordinates relevant in this article (Cartesian, polar and GGK) and we write down the String Newton-Cartan data for all of them.  In Appendix \ref{Appx_Cubic_Quartic} we list the cubic and quartic interactions obtained from the expansion of the action in the large string tension and radius limits.

\section{Non-relativistic limit of AdS$_5\times$S$^5$ string theory}
\label{sec:non_rel_general}

Our starting point is to consider the usual relativistic string action in AdS$_5\times$S$^5$.  We shall focus only on the bosonic sector. The relativistic action is then 
\begin{equation}
\label{rel_action}
S = - \frac{T}{2} \int \dd^2 \sigma \,  \bigg(\gamma^{\alpha\beta} \partial_{\alpha} X^{\mu} \partial_{\beta} X^{\nu} g_{\mu\nu} + \varepsilon^{\alpha\beta}  \partial_{\alpha} X^{\mu} \partial_{\beta} X^{\nu} b_{\mu\nu} \bigg) \ , 
\end{equation}
where $T$ is the string tension, $\sigma^{\alpha} = (\tau, \sigma)$, with $\alpha = 0, 1$, are the string world-sheet coordinates, $\gamma^{\alpha\beta} \equiv \sqrt{-h} h^{\alpha\beta}$ is the Weyl invariant combination of the inverse world-sheet metric $h^{\alpha\beta}$ and $h =$ det$(h_{\alpha\beta})$.  $X^{\mu} = (t, z_i, \phi, y_i)$, where $i=1,..., 4$, are the string embedding coordinates, which are bosonic fields depending on $(\tau, \sigma)$.  $b_{\mu\nu}$ is a \emph{closed} Kalb-Ramond field ($\dd b = 0$), which is left generic for the moment, and it will be fine-tuned to a critical value when taking the non-relativistic limit.  $g_{\mu\nu}$ is the AdS$_5\times$S$^5$ metric, which we choose to describe using Cartesian global coordinates,
\begin{equation}
\label{metric_cartesian}
g_{\mu\nu} \dd X^{\mu} \dd X^{\nu} = - \bigg(\frac{1+ \frac{z^2}{4 R^2}}{1-\frac{z^2}{4 R^2}}\bigg)^2 \dd t^2 + \frac{1}{(1-\frac{z^2}{4 R^2})^2} \dd z_i \dd z_i + \bigg(\frac{1 - \frac{y^2}{4 R^2}}{1 + \frac{y^2}{4 R^2}}\bigg)^2\dd \phi^2 + \frac{1}{(1+\frac{y^2}{4 R^2})^2} \dd y_i \dd y_i \ ,
\end{equation}
where $R$ is the common radius of AdS$_5$ and S$^5$. We use the short-hand notation $z^2\equiv z_i z^i$, where $z^i = z_i$, and the same for the $y$ coordinates.  In the limit $R \rightarrow \infty$ the action (\ref{rel_action}) describes strings moving in flat space. 

This relativistic action has a $\mathfrak{so}(4,2) \oplus \mathfrak{so}(6)$ global symmetry algebra, and the spacetime is described by a Riemannian metric which has local Lorentz invariance.  
To obtain the analogue in Cartesian coordinates of the non-relativistic AdS$_5\times$S$^5$ action found in \cite{Gomis:2005pg},  we need to contract $\mathfrak{so}(4,2) \oplus \mathfrak{so}(6)$ as described in Appendix \ref{Appx_NCdata}, which takes the $\mathfrak{so}(4,2)$ algebra to the 5d Newton-Hooke algebra.  

Rescaling the generators of $\mathfrak{so}(4,2) \oplus \mathfrak{so}(6)$ with a parameter $\omega$ induces a dual rescaling of coordinates,  via the coset representative of Cartesian coordinates given in eqn. (\ref{cartesian_coset}).  The induced rescaling of coordinates is the following
\begin{equation}
t \rightarrow t \ ,  \qquad 
z_1 \rightarrow z_1 \ ,  \qquad
 z_m \rightarrow \frac{1}{\omega} z_m \ , \qquad
 \phi \rightarrow \frac{1}{\omega} \phi \ ,  \qquad
 y_i \rightarrow \frac{1}{\omega} y_i \ , 
\end{equation}
where $m = 2, 3,4$. It is also convenient to rescale the string tension as follows
\begin{equation}
T \rightarrow \omega^2 T \ , 
\end{equation}
and absorb this $\omega^2$ factor inside the spacetime metric, namely
\begin{equation}
G_{\mu\nu} \equiv \omega^2 g_{\mu\nu} = \hat{E}_{\mu}{}^{\hat{A}} \hat{E}_{\nu}{}^{\hat{B}} \hat{\eta}_{\hat{A}\hat{B}} \ , 
\end{equation} 
where $\hat{\eta}_{\hat{A}\hat{B}} =$ diag$(-1, 1, ..., 1)$. The coordinate rescaling implies that the relativistic vielbein expands as
\begin{eqnarray}
\label{vielbeine_exp}
\hat{E}_{\mu}{}^A = \omega \tau_{\mu}{}^A + \frac{1}{\omega} m_{\mu}{}^A + \mathcal{O}(\omega^{-2})\ , \qquad\qquad
\hat{E}_{\mu}{}^a = e_{\mu}{}^a + \mathcal{O}(\omega^{-1}) \ , 
\end{eqnarray}
where $\hat{A} = 0, ... , d-1$ is decomposed into $A= 0, 1$, which are the \emph{longitudinal} directions in tangent space, and into $a = 2, ..., d-1$, which are the \emph{transverse} directions in tangent space. 
In our case, it turns out to be\footnote{In this article we trade some abuse of notation for some convenience by writing $\tau_\mu{}^A$ and $e_\mu{}^a$ as $10\times 10$ matrices, although they are $10\times 2$ and $10 \times 8$ matrices respectively.}
\begin{eqnarray}
\label{NR_vielbeine}
\notag
\tau_{\mu}{}^A &=& \text{diag}\bigg(-\frac{1+ (\frac{z_1}{2R})^2}{1- (\frac{z_1}{2R})^2},   \frac{1}{1 - (\frac{z_1}{2R})^2}, 0, ... , 0 \bigg) \ , \\
m_{\mu}{}^A &=& \text{diag}\bigg(-\frac{z_m z_m}{2R^2(1-(\frac{z_1}{2R})^2)},  \frac{z_m z_m}{4R^2(1-(\frac{z_1}{2R})^2)}, 0, ... , 0 \bigg) \ , \\
\notag
e_{\mu}{}^a &=&\text{diag} \left(0, 0, \frac{1}{1 - (\frac{z_1}{2R})^2}, \frac{1}{1 - (\frac{z_1}{2R})^2}, \frac{1}{1 - (\frac{z_1}{2R})^2}, 1,1,1,1,1 \right) \ .
\end{eqnarray}
Substituting the above expansion inside the relativistic action, we get
\begin{equation}
\label{S_NR_div}
S = - \frac{T}{2} \int \dd^2 \sigma \, \gamma^{\alpha\beta} \bigg( \omega^2 \partial_{\alpha} X^{\mu} \partial_{\beta} X^{\nu} \tau_{\mu\nu} + \partial_{\alpha} X^{\mu} \partial_{\beta} X^{\nu} H_{\mu\nu} + \varepsilon^{\alpha\beta}  \partial_{\alpha} X^{\mu} \partial_{\beta} X^{\nu} B_{\mu\nu}\bigg) + \mathcal{O}(\omega^{-2}) \ , 
\end{equation} 
where we introduced 
\begin{equation}
\tau_{\mu\nu} \equiv \tau_{\mu}{}^A \tau_{\nu}{}^B \tilde{\eta}_{AB} \ , \qquad\qquad
H_{\mu\nu} \equiv e_{\mu}{}^a e_{\nu}{}^b \tilde{\delta}_{ab} + \bigg(\tau_{\mu}{}^A m_{\nu}{}^B + \tau_{\nu}{}^A m_{\mu}{}^B \bigg) \tilde{\eta}_{AB} \ , 
\end{equation}
where $\tilde{\eta}_{AB} =$ diag$(-1, 1, 0, ..., 0)$, $\tilde{\delta}_{ab} =$ diag$(0,0,1,...,1)$. $\tau_{\mu\nu}$ is called the \emph{longitudinal} metric, while $H_{\mu\nu}$ is the \emph{boost invariant} two-tensor \cite{Bergshoeff:2018yvt}. 

In the contraction limit where $\omega \rightarrow \infty$, the term in the action containing $\tau_{\mu\nu}$ diverges. This is cured by fine-tuning the closed B-field to a critical value, 
\begin{equation}
B_{\mu\nu} \equiv \omega^2 b_{\mu\nu} = \omega^2 \tau_{\mu}{}^A \tau_{\nu}{}^B \varepsilon_{AB} \ , 
\end{equation}
where we have included in $B_{\mu\nu}$ the $\omega^2$ factor from the string tension redefinition. 
The convention used for the Levi-Civita symbol is $\varepsilon^{01} = + 1$ both for $\varepsilon^{\alpha\beta}$ and $\varepsilon^{AB}$.  By doing this, we get 
\begin{equation}
S  = - \frac{T}{2} \int \dd^2 \sigma \, \gamma^{\alpha\beta}\partial_{\alpha} X^{\mu} \partial_{\beta} X^{\nu} H_{\mu\nu} +  S_{div} + \mathcal{O}(\omega^{-2}) \ , 
\end{equation}
where now the divergent part nicely recasts into a Lorentz square 
\begin{equation}
\label{S_div_square}
S_{div} = - \frac{\omega^2 \, T}{2} \int \dd^2 \sigma \, \gamma^{00} \mathcal{F}^A \mathcal{F}^B \eta_{AB} \ , 
\end{equation}
where
\begin{equation}
 \mathcal{F}^A = \tau_{\mu}{}^A \partial_0 X^{\mu} - \frac{1}{\gamma_{11}} \varepsilon^{AB} \eta_{BC} \tau_{\mu}{}^C \partial_1 X^{\mu} - \frac{\gamma_{01}}{\gamma_{11}} \tau_{\mu}{}^A \partial_1 X^{\mu} \ . 
\end{equation}
The divergent action written in the form (\ref{S_div_square}) is suitable to be rewritten in terms of Lagrange multipliers $\lambda_{A}$ as 
\begin{equation}
S_{div} = - \frac{T}{2} \int \dd^2 \sigma \bigg( \lambda_A \mathcal{F}^A + \frac{1}{4 \omega^2 \gamma^{00}} \lambda_A \lambda^A \bigg) \ ,
\end{equation}
which is equivalent to (\ref{S_div_square}) by using the equations of motion for $\lambda_A$.  In this way the divergent action is just the sum of a finite and a subleading term in the limit $\omega\rightarrow \infty$. 

Finally, we take the limit $\omega\rightarrow \infty$ and we remain with the non-relativistic action
\begin{equation}
\label{NR_action}
S^{NR} = \lim_{\omega\rightarrow \infty} S =- \frac{T}{2} \int \dd^2 \sigma \, \bigg( \gamma^{\alpha\beta}\partial_{\alpha} X^{\mu} \partial_{\beta} X^{\nu} H_{\mu\nu} + \lambda_A \mathcal{F}^A \bigg) \ . 
\end{equation}
At this stage, one may want to integrate out the Lagrange multipliers. By imposing the equations of motion for the Lagrange multipliers one gets that $h_{\alpha\beta}$ is solved in terms of $\tau_{\alpha\beta}$ as follows
\begin{equation}
h_{\alpha\beta} \sim \tau_{\alpha\beta} \equiv \tau_{\mu\nu} \partial_{\alpha} X^{\mu}\partial_{\beta} X^{\nu} , 
\end{equation}
where $\sim$ means that $h_{\alpha\beta}$ is identified with the pull-back of the $\tau_{\mu\nu}$ metric up to a conformal factor. This implies that 
$\gamma^{\alpha\beta} = \sqrt{-\tau} \tau^{\alpha\beta}$, and the action (\ref{NR_action}) turns into the Nambu-Goto form
\begin{equation}
S^{NR}_{\footnotesize\text{NG}} = - \frac{T}{2} \int \dd^2 \sigma \,  \sqrt{-\tau}\tau^{\alpha\beta}\partial_{\alpha} X^{\mu} \partial_{\beta} X^{\nu} H_{\mu\nu}  \ . 
\end{equation}
Since in AdS$_5\times$S$^5$ the $\tau_{\mu\nu}$ metric describes an AdS$_2$ geometry,  in \cite{Gomis:2005pg} the authors imposed static gauge in order to obtain a 2d theory of free massive and massless bosons propagating in AdS$_2$. However, we will not follow this approach here. 

The non-relativistic action (\ref{NR_action}) describes strings propagating in a background which is not Riemannian, but String Newton-Cartan instead. String Newton-Cartan geometry is specified by the set of vielbein $\{ \tau_{\mu}{}^A , m_{\mu}{}^A, e_{\mu}{}^a\}$, and their local symmetry is the \emph{String Newton-Cartan algebra}\footnote{The String Newton-Cartan algebra is a particular non-central extension of the string Galilei algebra.} as amply discussed in \cite{Bergshoeff:2019pij}.  It is interesting to study physical properties of this theory as a theory standing on its own,  regardless of the limiting procedure to derive it, and in section \ref{sec:lighcone} we discuss light-cone gauge fixing for the non-relativistic AdS$_5\times$S$^5$ action in the large string tension limit.

\subsection{Change of coordinates}

The non-relativistic action shown in the previous section was obtained in a particular choice of coordinates, namely the Cartesian. The procedure was to fix our set of coordinates for the relativistic action in AdS$_5\times$S$^5$, apply the rescaling rules of coordinates induced by the contraction of $\mathfrak{so}(4,2)$ to the 5d Newton-Hooke algebra which singles out an AdS$_2$ subspace inside AdS$_5$, and read off the String Newton-Cartan data from the expansion of the relativistic vielbein. It is natural to wonder what happens if one derives the non-relativistic action in a different set of coordinates. In general, different sets of coordinates give rise to inequivalent non-relativistic actions.

The way to verify whether two non-relativistic actions written in different coordinates are equivalent, is to check if the two set of SNC data $\{ \tau_{\mu}{}^A, m_{\mu}{}^A, e_{\mu}{}^a \}$ belong to the same equivalence class. We say that two sets of SNC data are equivalent if they are related by an SNC gauge transformation.
The infinitesimal SNC gauge transformations are
\begin{align}
     \delta \tau_{\mu}{}^A &= \Lambda \,\varepsilon^A{}_B \tau_{\mu}{}^B \ ,  \\
     \delta e_\mu{}^{a} &= - \Lambda_A{}^{a} \tau_\mu{}^A + \Lambda^{a}{}_{b} \, e_\mu{}^{b}\ , \\
     \delta m_\mu{}^A  &= \partial_\mu \sigma^A - \varepsilon^A{}_B \sigma^B \Omega_\mu + \Lambda \, \varepsilon^A{}_B m_\mu{}^B + \Lambda^{A}{}_a e_{\mu}{}^a - \tau_\mu{}^B \sigma^A{}_B \ ,
\end{align}
where $\Lambda, \Lambda^{Aa}, \Lambda^{ab}, \sigma^A, \sigma^{AB}$, with $\sigma^A{}_A = 0$, are infinitesimal parameters associated with the SNC gauge transformations. We refer to \cite{Bergshoeff:2019pij} for further detail. To compare different sets of SNC data, infinitesimal SNC transformations are usually not enough, as the fields can differ by a finite quantity. For instance, $m_{\mu}{}^A$ in polar coordinates vanishes, whereas in Cartesian coordinates it has a finite non-zero value. 
In general, one would need a \emph{finite} SNC gauge transformation to relate the two sets of SNC data. Although we should be able to obtain them from the exponentiation of the infinitesimal transformations, this process can be quite involved. 
%In general, one should be able to find a \emph{finite} SNC gauge transformation to relate the two sets of SNC data. However, the specific form of these finite transformations is still unknown.
    
If the transformation that relates two coordinate systems is analytic in the $1/\omega$, we expect that the change of coordinates commutes with the process of taking the NR limit, and therefore the two sets of SNC data are equivalent in the sense described above.

\section{Uniform light-cone gauge in the large $R$ and $T$ limit}
\label{sec:lighcone}
In this section, we shall fix uniform light-cone gauge for the bosonic sector of non-relativistic AdS$_5\times$S$^5$ string theory \cite{Gomis:2005pg}. We consider the non-relativistic action (\ref{NR_action}) derived from Cartesian global coordinates, where the String Newton-Cartan vielbeine are given in (\ref{NR_vielbeine}).  The boost invariant tensor $H_{\mu\nu}$ reads
\begin{equation}
H_{\mu\nu} \dd X^{\mu} \dd X^{\nu} =  H_{tt} \dd t^2 + H_{\phi\phi} \dd \phi^2 + H_{IJ} \dd X^I \dd X^J \ , 
\end{equation}
where
\begin{eqnarray}
\notag
H_{tt} &=& - \frac{(1+ (\frac{z_1}{2 R})^2)z_m z_m}{R^2(1- (\frac{z_1}{2 R})^2)^3} \ , \qquad\qquad
H_{\phi\phi} = 1 \, , \\
H_{IJ} \dd X^I \dd X^J &=&  \frac{z_m z_m}{2 R^2(1- (\frac{z_1}{2 R})^2)^3}\,  \dd z_1^2 
+ \frac{1}{(1- (\frac{z_1}{2 R})^2)^2} \, \dd z_m \dd z_m 
+ \dd y_i \dd y_i \, ,
\end{eqnarray}
and we remind that the non-zero components of the longitudinal vielbein $\tau_{\mu}{}^A$ are
\begin{equation}
\tau_t{}^0 = -\frac{1+ (\frac{z_1}{2R})^2}{1- (\frac{z_1}{2R})^2} \ , \qquad\qquad
\tau_{z_1}{}^1 =\frac{1}{1 - (\frac{z_1}{2R})^2} \ . \label{tauvielbein}
\end{equation}

Here, we denoted by $X^I \equiv (z_1, z_m,  y_i)$ the coordinates transverse to the $(t, \phi)$-coordinates, which will be used to construct light-cone coordinates. We should not confuse $X^I$, which are coordinates trasverse to the light-cone, with $\{ t, z_1 \}$ and $\{ z_m, \phi, y_i \}$, which are non-relativistic longitudinal and transverse coordinates respectively. The two vector fields $\partial_t$ and $\partial_{\phi}$ are isometries for the String Newton-Cartan vielbeine (\ref{NR_vielbeine}), and therefore they leave the non-relativistic action invariant. In the limit $R\rightarrow \infty$ one recovers the string Newton-Cartan data for non-relativistic flat space derived from Cartesian coordinates.

The range of $\sigma$ is $0\leq \sigma\leq 2\pi$, and the string embedding coordinates are assumed to be periodic in $\sigma$, except for $z_1$, which is assumed to be anti-periodic.   In our analysis, we shall assume that the string does not wrap the $\phi$-direction.

To begin, we shall write the action (\ref{NR_action}) in the first order formalism. The momenta are defined as 
\begin{equation}
\label{def_momenta}
p_{\mu} \equiv \frac{\delta S^{NR}}{\delta \dot{X}^{\mu}} = - T \gamma^{0\alpha} \partial_{\alpha} X^{\nu} H_{\mu\nu} - \frac{T}{2} \lambda_A \tau_{\mu}{}^A \ , 
\end{equation}
where we use the notation $\dot{X}^{\mu} \equiv \partial_{\tau} X^{\mu}$ and $X^{\mu \, '} \equiv \partial_{\sigma} X^{\mu}$. The action then takes the form
\begin{equation}
\label{first_NR_action}
S^{NR} = \int \dd \tau \int_0^{2 \pi} \dd \sigma \, \bigg( p_{\mu} \dot{X}^{\mu} + \frac{\gamma^{01}}{\gamma^{00}} C_1 + \frac{1}{2 T \gamma^{00}} C_2 \bigg) \ , 
\end{equation}
where
\begin{equation}
C_1 = p_{\mu} X^{\mu \, '} \ , 
\end{equation}
and 
\begin{eqnarray}
\notag
C_2 &=& H^{\mu\nu} p_{\mu}p_{\nu} + T^2 H_{\mu\nu} X^{\mu \, '}X^{\nu \, '} + T \lambda_A \tau_{\mu}{}^A p_{\nu} H^{\mu\nu} \\
&+& \frac{T^2}{4} \lambda_A \lambda_B \tau_{\mu}{}^A \tau_{\nu}{}^B H^{\mu\nu}
- T^2 \lambda_A \varepsilon^{AB} \eta_{BC} \tau_{\mu}{}^C X^{\mu \, '} \ ,
\end{eqnarray}
are combinations of the Virasoro constraints.\footnote{Precisely, $C_1$ and $C_2$ are linear combinations of the world-sheet stress-energy tensor
\begin{equation}
\notag
C_1 =- h \bigg(  \gamma^{01} T^{00} - \gamma^{00} T^{01} \bigg) \ , \qquad\qquad
C_2 = 2 h T\,  T^{00} \ , \qquad\qquad  T^{\alpha\beta} \equiv - \frac{2}{\sqrt{h}} \frac{\delta S^{NR}}{\delta h_{\alpha\beta}}\ .
\end{equation}}
Here $H^{\mu\nu}$ is the usual inverse of $H_{\mu\nu}$, i.e. $H^{\mu\rho}H_{\rho \nu}=\delta^\mu_\nu$.
\footnote{Instead, we could have used a non-relativistic action where the St\"uckelberg symmetry has been fixed, so that it is described by a degenerate tensor $H^{\perp}_{\mu\nu} = e_{\mu}{}^a e_{\nu}{}^b \tilde{\delta}_{ab}$ and a non-relativistic B-field $B^{\text{NR}}_{\mu\nu} = \varepsilon_{AB} \tau_{\mu}{}^A m_{\nu}{}^B$. In this case, we should be able to write the first-order action using the \emph{projective inverse}, i.e. $H^{\mu\rho}H^{\perp}_{\rho \nu}= (\delta^\mu_\nu - \tau^{\mu}{}_A \tau_{\nu}{}^A)$. The final result should be similar to ours, but with the additional new term coming from the non-relativistic B-field.}  
We refer to \cite{Kluson:2018grx} for its expression in terms of the Newton-Cartan data. It is interesting to note that although the non-relativistic action apparently has a different structure from the relativistic one, at the end it is still possible to bring it into this form, which indicates that our action still comes from a Hamiltonian that is the sum of two first class constraints.

The action (\ref{first_NR_action}) is invariant under a transformation that changes the sign of the transverse coordinates. However, the transformation that changes the sign of $z_1$, which was a symmetry of the original AdS$_5\times$S$^5$ appears to be broken at first sight. If we look a bit more in detail, we will notice that the transformation $z_1 \to -z_1$ is clearly a symmetry of the action in the Nambu-Goto formalism. Furthermore, the action in the Polyakov formalism written by using the zwiebein of the world-sheet metric has a $z_1$ parity symmetry, which requires at the same time to act also on the Lagrange multipliers and on the zweibein, see \cite{Fontanella:2021btt}.

Now we wonder if there is a way to make the $z_1 \to -z_1$ symmetry manifest on the action (\ref{first_NR_action}). The answer is positive, and it is given in terms of a parity transformation on flat indices
\begin{eqnarray}
    \lambda_1 \rightarrow -\lambda_1 \ , \qquad
    \varepsilon^{AB} \rightarrow -\varepsilon^{AB}  \ , \qquad
    \tau_{z_1}{}^1 \rightarrow -  \tau_{z_1}{}^1 \ . 
\end{eqnarray}
Here $\varepsilon^{AB}$ changes sign because the parity transformation changes the orientation of the flat internal space. Thanks to the specific form of our $\tau$ vielbein (\ref{tauvielbein}), namely that it is diagonal and quadratic in $z_1$, under the change of sign $z_1 \to -z_1$ it transforms as
\begin{eqnarray}
    \tau_{z_1}{}^1 \partial_{\alpha} z_1 \to -  \tau_{z_1}{}^1 \partial_{\alpha} z_1 \ .
\end{eqnarray}
Therefore, the previous symmetry of the action can also be expressed as 
\begin{eqnarray}
\label{Z2_symmetry}
    z_1\rightarrow -z_1 \ , \qquad
    p_{z_1} \rightarrow - p_{z_1} \ , \qquad
    \lambda_1 \rightarrow -\lambda_1 \ , \qquad
    \varepsilon^{AB} \rightarrow -\varepsilon^{AB} \ .
\end{eqnarray}
Notice that we had to change the sign of $p_{z_1}$, as it is associated to the time derivative of $z_1$ through (\ref{def_momenta}).

We will make use of this symmetry in the following section to write a classical solution of the equations of motion associated to (\ref{first_NR_action}) with twisted periodic boundary condition on the $z_1$ coordinate.

To proceed further in our computations, we need to fix the light-cone gauge. For this, it is necessary to introduce light-cone coordinates
\begin{eqnarray}
\notag
X_+ &=& (1-a) t + a \phi \ , \qquad\ \ 
X_- = \phi - t \ , \qquad \\
p_+ &=& (1-a) p_{\phi} - a p_t \ , \qquad
p_- = p_{\phi} + p_t \ , 
\end{eqnarray}
where $a$ is a gauge freedom that we will leave unfixed\footnote{The purpose of the parameter $a$, which was first introduced in \cite{Arutyunov:2006gs}, is to check the correctness of the procedure, as observables, like the spectrum, have to be independent of it.}. We fix uniform light-cone gauge by imposing the following conditions
\begin{equation}
\label{lightcone_gauge}
X_+ = \tau \ , \qquad\qquad p_+ = K \, T \ ,
\end{equation}
where $K$ is a constant, fixed as follows
\begin{equation}
K = a \kappa - |\kappa|\sqrt{1-2a+2a^2} \ ,
\end{equation}
where $a$ is the gauge parameter and $\kappa$ is a \emph{non-zero} real number, entering the twisted BMN-like solution, which will be presented later in (\ref{winding_eqn}). This constant is real because the argument inside the square root is never negative for $0\leq a \leq 1$. This choice of $K$ avoids the appearance in the action of imaginary units, and it will canonically normalise, up to an overall factor, the quadratic free-fields Hamiltonian. 

The light-cone momentum $P_+$ defined as 
\begin{equation}
P_+ \equiv \int_0^{2\pi} \dd \sigma \, p_+ \ , 
\end{equation}
is fixed in terms of the string tension as
\begin{equation}
P_+ = 2\pi K \,T \ , 
\end{equation}
which is obtained by integrating over $\sigma$ the light-cone gauge condition (\ref{lightcone_gauge}).  We are interested in studying the non-relativistic action in the limit where $T\rightarrow \infty$ and $P_+ \rightarrow \infty$, keeping $P_+ / T$ fixed to $2\pi K$. 

To proceed, we integrate the Lagrange multipliers out.  To do so, we point out that the conjugate momenta to $\lambda_A$ is identically zero, since in the non-relativistic action (\ref{NR_action}) there is no $\dot{\lambda}_A$ appearing. Therefore, we have identically that 
\begin{equation}
p_{\lambda_A} = 0 \ .
\end{equation}
This condition must be preserved in the world-sheet time evolution, which means that
\begin{equation}
\partial_{\tau} p_{\lambda_A} = \{ p_{\lambda_A}, \mathcal{H}\} \approx 0 \ ,
\end{equation}
where the symbol $\approx$ indicates equality up to terms proportional to the constraint. This gives two independent equations that can be solved for $(\lambda_0, \lambda_1)$ in terms of coordinates and momenta.  By plugging their solution back into (\ref{first_NR_action}) one gets an action fully independent of the Lagrange multipliers. 

Next,  we need to impose the Virasoro constraints $C_1 = C_2  \approx  0$.  The first Virasoro constraint, after fixing light-cone gauge, can be solved in terms of $X'_-$
\begin{equation}
C_1 = p_+ X'_- + p_I X^{I \, '}  \approx  0 \ ,  \qquad \Longrightarrow \qquad X'_- = - \frac{1}{K\, T}\, p_I X^{I \, '} \ . 
\end{equation}
The second Virasoro constraint $C_2  \approx  0$, in the light-cone gauge and after solving the first Virasoro constraint in terms of $X'_-$,  is an equation that only depends on $p_-, X^I$, $X^{I\, '}$, $p_I$.  In particular, this is a quadratic equation in $p_-$ which can be solved in terms of $X^I$, $X^{I\, '}$, $p_I$.  
In this way, the action (\ref{first_NR_action}) becomes 
\begin{equation}
S^{NR} = \int \dd^2 \sigma \, (p_I \dot{X}^I - \mathcal{H} ) \ , 
\end{equation}
where
\begin{equation}
\label{Ham_p-}
 \mathcal{H} = - p_- (X^I, X^{I\, '}, p_I) \ , 
\end{equation}
and where we neglected the total derivative $\dot{X}_-$.

\subsection{The expansion}

\begin{figure}[t]
\begin{center}
 \includegraphics[scale=1]{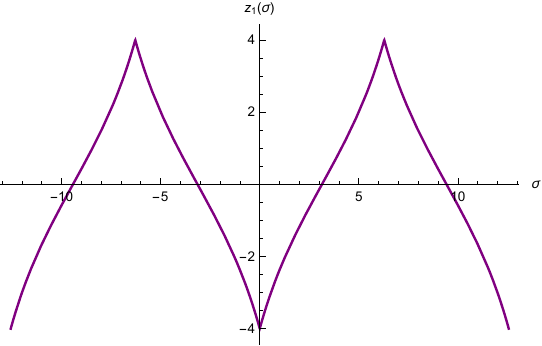}
\end{center}
\caption{$z_1$ profile of the twisted periodic BMN-like solution for values of the parameters $\kappa = 1$ and $R=2$. Courtesy of  \cite{Fontanella:2021btt}.} 
\label{z1gluing}
\end{figure}

At this stage, we fixed light-cone gauge and all constraints have been solved, therefore we remained with an action that depends only on the minimal physical degrees of freedom of the theory. The next step is to expand our action around a classical solution, which we choose to be the twisted periodic BMN-like solution \cite{Fontanella:2021btt}. In the fundamental domain $\sigma \in [0, 2 \pi]$, the solution is
\begin{eqnarray}
\label{classical_sol}
    t= \kappa \tau \ , \quad
    z_1 = 2R \tan \left[ -\frac{\kappa}{2 R} (\sigma-\pi) \right] \ , \quad
    \phi = \omega \tau \ , \quad
    \lambda_0 = \frac{\omega}{\kappa}\cos \left[ \frac{\kappa}{R} (\sigma-\pi) \right] \ ,
\end{eqnarray}
where $\kappa, \omega$ are two unconstrained real parameters.
Outside the $\sigma$ fundamental domain, this solution is defined via a gluing procedure, such that the embedding coordinates are $2\pi$-periodic in $\sigma$, except of $z_1$ and $\lambda_1$ which are anti-periodic, as allowed by the $\mathbb{Z}_2$ symmetry (\ref{Z2_symmetry}). The profile for $z_1$ is shown in figure \ref{z1gluing}, and due to the gluing procedure, the derivative of $z_1$ is not defined at $\sigma = 2\pi n$ for $n \in \mathbb{Z}$. However, it is still possible to assign a value for the derivative of $z_1$ in these points by using a small parameter $\varepsilon$ prescription, namely $z_1' (2 \pi n) \equiv \lim_{\varepsilon \rightarrow 0^+} z_1'(2 \pi n  + \varepsilon)$, where $\varepsilon$ is here conventionally assumed to be a small positive quantity. With this prescription, the derivative of $z_1$ in $2\pi n$ has alternating sign. 
%%%%%% Figure gluing
The solution (\ref{classical_sol}) was found in conformal gauge, but it was checked in \cite{Fontanella:2021btt} that it also satisfies the 
equations of motion in light-cone gauge with all constraints implemented, as relevant in our case.

It is important to note that, as it was shown in \cite{Fontanella:2021btt}, solving the equations of motion imposed by the Lagrange multipliers implies that both non-relativistic longitudinal coordinates $t$ and $z_1$ must be non-zero. Therefore, by requiring that the vacuum has $t= \kappa \tau$, it requires for consistency that $z_1$ behaves like (\ref{classical_sol}). This has a remarkable consequence on the expansion of the non-relativistic action around this solution, as $z_1$ is not an isometry, and terms depending on $\sigma$ will appear in the expanded action. However, as we shall see, these $\sigma$ terms may be regarded as corrections to a free-field theory at large $R$.

We are now ready to expand the light-cone gauge fixed action around the twisted periodic BMN-like vacuum. First, we fluctuate the embedding coordinates around the classical solution $X^I = X^I_{\text{cl}} + \tilde{X}^I$, which amounts to defining
\begin{equation}
\label{winding_eqn}
z_1 = z_{\text{cl}} +  \tilde{z} \ , \qquad\qquad
z_{\text{cl}} = 2R \tan \left[ -\frac{\kappa}{2 R} (\sigma-\pi) \right] \ , 
\end{equation}
where $\tilde{z} \equiv \tilde{z}(\tau, \sigma)$ is a fluctuation and assumed to satisfy twisted periodic boundary conditions, i.e. $\tilde{z}(\tau, \sigma + 2\pi) = - \tilde{z}(\tau, \sigma)$. For the remaining coordinates appearing in the gauge fixed Lagrangian, the classical solution is identically zero, and therefore we omit the tilde notation on their fluctuations.    
Next, we rescale fields and momenta with the following canonical transformation\footnote{There is also a \emph{non-canonical} expansion that gives same results as the canonical one.  In this case, one needs to fix the light-cone gauge condition $p_+ = K$ instead of (\ref{lightcone_gauge}),  to scale $\sigma\rightarrow T \sigma$ and to rescale both coordinates and momenta with $1/ \sqrt{T}$. The non-canonical expansion has the advantage of being more efficient computationally speaking.  We thank G.  Arutyunov for discussion.} 
\begin{eqnarray}
\notag
&&\tilde{z} \rightarrow \frac{1}{\sqrt{T}} \tilde{z} \ , \qquad\quad\ 
z_m \rightarrow \frac{1}{\sqrt{T}} z_m \ , \qquad\quad
y_i \rightarrow \frac{1}{\sqrt{T}} y_i \ , \\
&&p_{z_1} \rightarrow \sqrt{T} p_{\tz} \ , \qquad
p_{z_m} \rightarrow \sqrt{T} p_{z_m} \ , \qquad
p_{y_i} \rightarrow \sqrt{T} p_{y_i} \ .
\end{eqnarray}
Notice that the classical momentum associated with $z_{\text{cl}}$ is identically zero.
For the purpose of the expansion, we choose $\kappa = -1$ and $a=1$, for which it corresponds $K = -2$ from (\ref{lightcone_gauge}). 
In the large $T$ and $R$ limit the gauge fixed non-relativistic action expands as
\begin{equation}
\label{S_expan}
S^{NR}_{gf} = \int \dd \tau \dd \sigma \bigg( \mathcal{L}_2 + \frac{1}{R^2}\hat{\mathcal{L}}_2+  \frac{1}{\sqrt{T}} \mathcal{L}_3 + \frac{1}{R^2\,\sqrt{T}}\hat{\mathcal{L}}_3 + \frac{1}{T} \mathcal{L}_4 + \frac{1}{R^2\,T}\hat{\mathcal{L}}_4 + ....\bigg) \ , 
\end{equation}
where we omitted the constant term and the total derivative term linear in the fields, which are irrelevant for the dynamics. After rescaling the world-sheet time $\tau \rightarrow 2 \tau$ in order to make the overall coefficient of the quadratic Hamiltonian canonical\footnote{The quadratic Hamiltonian appears in the expansion with an overall factor of $1/4$, instead of the canonical $1/2$.}, we find that the leading term $\mathcal{L}_2$ describes 8 free massless fields in 2d flat space,
\begin{equation}
\mathcal{L}_2 = p_I \dot{X}^I - \frac{1}{2}\bigg(p_{\tz}^2 + p_{z_m}^2 + p_{y_i}^2 + \tilde{z}'^2 + z_m'^2 + y_i'^2 \bigg) \ .
\end{equation}
The $\hat{\mathcal{L}}_2$ term is a radius correction and it is quadratic in the fields,
\begin{eqnarray}
\label{hatL2}
\notag
\hat{\mathcal{L}}_2 &=&-\frac{1}{4}\bigg[ 2 \tz^2 + 3 z_m^2 
+(\sigma-\pi) ^2 ( p_{\tz}^2- p_{z_m}^2 ) + 2s (\sigma-\pi) 
   \tz \tilde{z}'-(\sigma-\pi)^2 \tilde{z}'^2
   +(\sigma-\pi)^2 z_m'^2 \bigg]\ ,\\
\end{eqnarray}
where $s \equiv s(\sigma)$ is defined as $s = +1$ for $\sigma \in [0, 2\pi[$, and $s = -1$ for $\sigma = 2\pi$. The function $s$ appears to take into account the definition of derivative of $z_{\text{cl}}$ in the extrema of the fundamental domain, which flips sign as discussed above, and to keep track of the $\mathbb{Z}_2$ symmetry acting on a remnant of $\varepsilon^{AB}$. 
From the expression of $\hat{\mathcal{L}}_2$, we notice that $\tilde{z}$ and $z_m$ gain a mass term as a $1/R^2$ correction. 

The cubic and quartic Lagrangians are given in Appendix \ref{Appx_Cubic_Quartic}, and they do not show any pathological behaviour (e.g. non-locality, broken Hermiticity).  Moreover, the structure of all perturbative terms shows that the light-cone gauge preserves a manifest $SO(3) \times SO(4)$ symmetry.

Physical states should satisfy the level-matching condition, which is obtained by integrating out in $\sigma$ the linear Virasoro constraint $C_1  \approx  0$.  In our setting, where the string has no winding mode around $\phi$, the level-matching condition reads
\begin{equation}
p_{\text{ws}} =  \int_0^{2\pi} \dd \sigma \, p_{\tilde{z}} z_{\text{cl}}'\ , 
\end{equation} 
where
\begin{equation}
p_{\text{ws}} \equiv -\int_0^{2\pi} \dd \sigma ( p_{\tilde{z}} \tilde{z}' + p_{z_m}z_m' + p_{y_i}y_i' ) \ ,
\end{equation}
is the total world-sheet momentum of the string, and $z_{\text{cl}}$ is the classical solution around which the expansion in eqn. (\ref{winding_eqn}) takes place. Here we observe two things:
\begin{itemize}
\item the gauge fixed action is not invariant under shifts of $\sigma$, and therefore $p_{\text{ws}}$ is not a conserved charge. 
\item in the near-BMN limit of the relativistic theory, it was found that a winding mode around the $\phi$-direction makes $p_{\text{ws}}$ to be large \cite{Arutyunov:2004yx}.  This is in turn responsible of making the quadratic Hamiltonian to be large on physical states.  We remark that this does not happen in our setting since the level-matching condition implies that $p_{\text{ws}}$ is finite.  However if we were instead to turn on a winding mode around the $\phi$-direction, then we would find in the non-relativistic theory that $p_{\text{ws}}$ is large and the quadratic Hamiltonian is large on physical states,  like in the relativistic theory\footnote{We thank G.  Arutyunov for discussion.}.
\end{itemize}

We comment now on the choices of $\kappa$ and $a$.  We can identify three different scenarios, accordingly to the three choices of $a = 0, \frac{1}{2}, 1$.  
\begin{itemize}
\item  {\boldmath$a=0$}.  In this case one can derive an expansion similar to (\ref{S_expan}), but with the only difference that the sign of the quadratic Hamiltonian is governed by the sign of $\kappa$.  Only for positive $\kappa$ one has a positive definite quadratic Hamiltonian\footnote{This is because when $a=0$ the Hamiltonian obtained by solving the Virasoro constraints is schematically of the form 
\begin{equation}
\notag
\mathcal{H} = A x \pm \sqrt{B x + C}\, 
\end{equation}
where $x$ is the Lagrange multiplier and $A, B, C$ are generic expressions depending on the fields.  To integrate out the Lagrange multipliers, one has to solve this type of equation 
\begin{equation}
\notag
\frac{\dd \mathcal{H} }{\dd x} = A \pm \frac{B}{2 \sqrt{B x + C}} \overset{!}{=} 0 \ , \qquad \Longrightarrow \qquad
\mathcal{H} = \frac{A}{B} \left[\left(\frac{B}{2A}\right)^2 - C \right] - \frac{B}{2 A} \ .
\end{equation} 
This result points out that $\mathcal{H}$ is clearly independent of the choice one can make between the two solutions of the second Virasoro constraint, once the Lagrange multipliers are integrated out.  This does not happen for the other choices of gauge parameter $a$, where one can always find one solution of the second Virasoro constraint that gives a positive definite quadratic Hamiltonian.}. 
\item {\boldmath$a=\frac{1}{2}$}.  In this case we obtain a similar expansion as in (\ref{S_expan}). The positive definite quadratic Hamiltonian exists for every choice of $\kappa$.  Computations are more involved in this gauge.
\item {\boldmath$a=1$}. In this case computations turn out to be simpler. The positive definite quadratic Hamiltonian exists for every choice of $\kappa$.  
\end{itemize}

Finally, we close this section by commenting on the expansion around different vacua. In particular, a first natural thing to do is to expand the action around the BMN-like solution with the periodic boundary condition. The $z_1$ profile of such solution is 
\begin{eqnarray}
\label{z1_periodic}
     z_1^{\footnotesize\text{periodic}\normalsize} = -2R \tan \left( \frac{n}{2} \sigma \right) \ , \qquad
     n \in \mathbb{Z} \ .
\end{eqnarray}
The difference with the twisted periodic BMN-like solution (\ref{classical_sol}) is that the $z_1$ coordinate of the latter one expands at large $R$ as
\begin{eqnarray}
\label{twisted_z1_large_R}
    z_1^{\footnotesize\text{twisted}\normalsize} = - \kappa (\sigma - \pi) + \mathcal{O}(R^{-2}) \ , 
\end{eqnarray}
i.e. it linearises in $\sigma$, whereas (\ref{z1_periodic}) retains the tangent behaviour at large $R$. The Hamiltonian quadratic in fluctuations is schematically of the form $(\partial X_{\text{fl}})^2 g^{\text{cl}}$, where $g^{\text{cl}}$ is a function of the background data $H_{\mu\nu}$ and $\tau_{\mu}{}^A$ evaluated on the classical solution. The classical solution introduces $\sigma$ terms via the $z_1$ coordinates, and $H_{\mu\nu}$ and $\tau_{\mu}{}^A$ depend on $z_1$ only via the combination $b_{\pm} \equiv 1 \pm (z^{\text{cl}}_1/ 2R)^2$. Then we need to ask what happens to $b_{\pm}$ at large $R$. For the twisted periodic solution which has expansion (\ref{twisted_z1_large_R}), we have that $b_{\pm} = 1 + \mathcal{O}(R^{-2})$, whereas for the periodic solution (\ref{z1_periodic}), we have  $b_{\pm} = 1 \pm \tan^2 \left(n \sigma /2 \right)$. In order to have a quadratic Hamiltonian describing free fields, only the $b_{\pm} \sim 1$ guarantees the absence of $\sigma$ terms, and therefore one needs to expand around the twisted periodic vacuum.

\subsection{Comments on finite $R$}
\label{sec:finiteR}

In the previous section we showed that if the radius $R$ is sufficiently large and the string expands around a classical solution whose $z_1$ coordinates linearises in $\sigma$ in the large radius expansion, the light-cone gauge fixed non-relativistic action admits a large string tension limit around the dynamics of free massless scalar fields.
The role of the large $R$ expansion is to render the $z_1$ coordinate an isometry in first approximation. 
The next question we want to ask is the following: \emph{what happens if we keep $R$ finite?}

If one tries to implement the procedure described in the previous section for finite $R$, one will find that the action does not expand around the dynamics of free particles, but instead it will expand around a more complicated quadratic action,  which is not free due to the introduction of $\sigma$ dependent terms, 
since $z_1$ is not an isometry for $H_{\mu\nu}$ and $\tau_{\mu}{}^A$.

The large $R$ limit has the property of taming the $\sigma$ dependent terms, in such a way that they disappear from the leading order action and only appear through subleading corrections.  The large $R$ limit has also the effect of making the $\tz$ and $z_m$ fields becoming massless, and their mass only appears as a correction in $1/R^2$. This is in agreement with the fact that non-relativistic string theory in flat spacetime (in static gauge) is just a theory of massless scalar fields in Mink$_2$.

One may wonder whether the problem of getting a complicated leading order action with $\sigma$ terms is just because of a bad choice of coordinates. To analyse further this question, first we remark that the following conditions are necessary in order to apply the light-cone gauge to our problem:
\begin{enumerate}
\item There must exist two coordinates, one timelike and one spacelike, which are isometries for $H_{\mu\nu}$ and $\tau_{\mu}{}^A$ (i.e.  the latter do not depend on those coordinates), such that they can be used to define light-cone directions $X_{\pm}$.
\label{time_space_like}
\item The two non-relativistic longitudinal directions must be coordinates describing an AdS$_2$ spacetime.  The timelike direction must be a non-relativistic longitudinal direction.
\label{AdS2}
\item The classical string must have winding around the non-relativistic longitudinal \emph{spatial} direction, for consistency with the Lagrange multipliers equations of motion. 
\label{winding}
\end{enumerate}
By keeping in mind the above conditions, we can notice the following. The two non-relativistic longitudinal directions cannot be chosen to be the timelike and spacelike isometries given in point (\ref{time_space_like}).  If this was possible,  it would imply that the longitudinal metric $\tau_{\mu\nu}$ describes the 2d Minkowski spacetime, and therefore (\ref{AdS2}) would not be fulfilled. 

Next, we consider the set of coordinates in AdS$_5\times$S$^5$ given in Appendix \ref{Appx_NCdata}, and we discuss their advantages and disadvantages regarding conditions (\ref{time_space_like})-(\ref{winding}). 

\vspace{2mm}
\emph{GGK coordinates.} In this set of coordinates, the longitudinal spatial coordinate $x_1$ is an isometry, and therefore it looks promising in terms of winding the string, since this will not generate any $\sigma$ dependent term. However, in agreement with the argument above, the timelike coordinate $x_0$ is not an isometry in this case, and hence this set of coordinates does not fulfil point (\ref{time_space_like}). In other words, if one tries to fix light-cone gauge in this set of coordinates, one will end up at leading order with a time-dependent Hamiltonian.   

\vspace{2mm}
\emph{Cartesian global coordinates.} In this set of global coordinates, all conditions (\ref{time_space_like})-(\ref{winding}) are satisfied\footnote{In this article, we considered only \emph{global} set of coordinates for AdS$_5\times$S$^5$. If one was allowed to consider also \emph{local} set of coordinates, then one could check what happens, for instance, in the Poincar\'e coordinates.}.  Since the (longitudinal) timelike direction is an isometry, the longitudinal spatial direction cannot be an isometry as well and one has to deal with the $\sigma$ terms generated by wrapping the longitudinal spatial direction, which we were able to shift them into large $R$ corrections. 

\vspace{2mm}
\emph{Polar global coordinates.} This set of coordinates has the same features as the Cartesian global coordinates, with the difference that $H_{\mu\nu}$ is not invertible, which nevertheless seems not to be a problem in order to apply the procedure described in this paper.

\section{Conclusions}
\label{sec:conclusions}

In this article, we studied the non-relativistic limit of the  AdS$_5\times$S$^5$ string theory action in different coordinates. At the relativistic level, there is no physical difference between actions written in different coordinate systems. However, the coordinate transformation might not commute with the process of computing the limit, leading to inequivalent theories. We expect that, if the transformation is analytic in $1/\omega$, there should be no problem. However, to truly check if that is the case, we would need to study if there is a finite SNC gauge transformation relating the two sets of SNC data.
Currently, only infinitesimal SNC gauge transformations are available in the literature. The finite SNC gauge transformations can be computed by exponentiating the infinitesimal ones, which is a task that we leave for future research.

In the second part of this article, we considered the non-relativistic AdS$_5\times$S$^5$ string action derived from Cartesian coordinates, and we fixed light-cone gauge. Then we studied the semiclassical expansion at large string tension of the gauge fixed action around the twisted periodic BMN-like solution found in \cite{Fontanella:2021btt}, and we showed the action expands about free fields. The $z_1$ direction of the classical solution depends non-trivially on $\sigma$, and since $z_1$ is not an isometry, we proposed in addition to take a large AdS radius expansion in order to shift the $\sigma$ terms into large radius corrections. We discussed that this procedure can be applied to the vacuum with twisted periodic boundary conditions, but not in the periodic case.  

The next step would be to consider the perturbative S-matrix associated with the expanded action. This step would require to take the so-called decompactification limit, where the topology of the world-sheet changes from being the one of the cylinder to the one of the infinitely extended 2d plane, where scattering states are prepared at the infinities. In this procedure, the $\sigma / R$ terms, which in the cylinder topology represent corrections at large $R$ to the free field dynamics, may contribute to the quadratic Lagrangian in the decompactification limit, as $\sigma$ is allowed to span the real line. It would be interesting to explore if there is any  procedure that allows to re-sum the $R$ corrections before taking the decompactification limit.

There are interesting future directions regarding the expansion of the action around the periodic BMN-like vacuum. As shown in \cite{Fontanella:2023men}, such vacuum arise from the non-relativistic limit of the non-compact folded string with zero spin. Such relativistic solution has infinite energy, but it was shown that the contribution from the closed critical B-field makes it finite. Because of its apparent infinite energy, such solution has not been properly studied in the past as a vacuum for the relativistic action. However, as now we know such infinity is precisely tamed by the critical closed B-field, one could try to expand the relativistic action around it, similarly to \cite{Frolov:2002av, Beccaria:2010ry}, and the non-relativistic action may also be similarly expanded around the periodic BMN-like vacuum. Although we showed the action does not expand around free fields, it might still be possible to compute quantum corrections to the vacuum classical energy.        

Other interesting, more general, questions regard the inclusion of  \emph{supersymmetry} into the procedure of light-cone gauge fixing and understanding what is the holographic dual field theory of non-relativistic AdS$_5\times$S$^5$ string theory. Answering the latter question requires defining what the \emph{boundary geometry} of a String Newton-Cartan manifold, which in general is still an open problem. For past work connected to this topic,  see  \cite{Sakaguchi:2007ba}.

\vskip 0.5cm
\noindent{\bf Acknowledgements} \vskip 0.1cm
\noindent 

\noindent 
We are in debt of gratitude to our collaborator A. Torrielli, who co-authored a previous version of this manuscript but decided to opt out in this new revised version. We acknowledge that many of the ideas presented in this work emerged during our collaboration.  
We thank G. Arutyunov, N. Obers and G. Oling for reading a first version of this manuscript and for providing very useful comments and insights. In addition, we thank A. Tseytlin and S. Frolov for useful discussions related to this work. AF also thanks S. van Tongeren for useful discussions and for a collaboration on related topics. AF has been supported by the German Research Foundation DFG via the Emmy Noether program ``Exact Results in Extended Holography''. JMNG and AT are supported by the EPSRC-SFI grant EP/S020888/1 \emph{Solving Spins and Strings}.
AF thanks Lia for her permanent support.

%%%%%%%%%%%% APPENDICIES %%%%%%%%%%%%%%%%%%%%%%

\setcounter{section}{0}
\setcounter{subsection}{0}
\setcounter{equation}{0}

\begin{appendices}

\section{String Newton-Cartan data}
\label{Appx_NCdata}

In this section we shall derive the set of Newton-Cartan vielbeine (or ``data'') $\{ \tau_{\mu}{}^A , m_{\mu}{}^A, e_{\mu}{}^a\}$ for various set of coordinates of AdS$_5\times$S$^5$. The idea consists of the following steps:
\begin{enumerate}
\item Choose a given coset representative $g \in SO(4,2)\times SO(6)$. This determines the choice of coordinates for AdS$_5\times$S$^5$. In what follows we will choose coset representatives corresponding to three sets of global coordinates: Cartesian, polar and GGK. 
\item Rescale the generators of the isometry algebra $\mathfrak{so}(4,2)\oplus\mathfrak{so}(6)$ by a parameter $\omega$. This rescaling is made in such a way that if one takes the limit $\omega$ to $\infty$, the $\mathfrak{so}(4,2)$ algebra contracts to the 5d Newton-Hooke algebra. In general there are several ways of making such contraction, which are all equivalent from the algebraic point of view. However not all of them can give a sensible physical result.  The physically sensible choice is the one that singles out an AdS$_2$ space inside AdS$_5$, where the $\mathfrak{sl}(2, \mathbb{R})$ symmetry preserved by the contraction limit can act on. 
\item Compute the left-invariant Maurer-Cartan 1-form $g^{-1} \dd g$.
\item Extract the Newton-Cartan data, i.e. the non-relativistic vielbeine, by expanding the relativistic vielbein in powers of $\omega$. 
\end{enumerate}

\subsection{$\mathfrak{so}(4,2)\oplus\mathfrak{so}(6)$ algebra contraction}

The rescaling of the $\mathfrak{so}(4,2)\oplus\mathfrak{so}(6)$ generators which makes the contraction of $\mathfrak{so}(4,2)$ to 5d Newton-Hooke is the following. We will follow the notation and convention used in \cite{Arutyunov:2009ga}. 
The generators of $\mathfrak{so}(4,2)$ are $m^{\hat{i}\hat{j}}$, with $\hat{i}, \hat{j}, ... = 0, ...,5$, where $0, 5$ describe the two time directions of the space where $\mathfrak{so}(4,2)$ is acting on, and they satisfy the algebra
\begin{equation}
[m^{\hat{i} \hat{j}} , m^{\hat{k}\hat{\ell}}] = \eta^{\hat{j}\hat{k}} m^{\hat{i}\hat{\ell}} - \eta^{\hat{i}\hat{k}}m^{\hat{j}\hat{\ell}} + \eta^{\hat{i}\hat{\ell}}m^{\hat{j}\hat{k}} - \eta^{\hat{j}\hat{\ell}} m^{\hat{i}\hat{k}} \, , 
\end{equation}
where $\eta =$ diag$(-1, 1, 1, 1, 1, -1)$.
One can identify the $\mathfrak{so}(4,1)$ subalgebra by choosing one particular time direction, e.g. $\hat{j}=5$, such that $m^{i5} \equiv P^i$ are the momentum generators on AdS$_5$, and $m^{ij} \equiv J^{ij}$ are the angular momentum generators spanning $\mathfrak{so}(4,1)$, where $i, j, ... = 0, ..., 4$. 

The generators of $\mathfrak{so}(6)$ are denoted by $n^{\hat{a}\hat{b}}$, with $\hat{a}, \hat{b}, ... = 1, ... , 6$, and they satisfy the algebra
\begin{equation}
[n^{\hat{a} \hat{b}} , n^{\hat{c}\hat{d}}] = \delta^{\hat{b}\hat{c}} n^{\hat{a}\hat{d}} - \delta^{\hat{a}\hat{c}}n^{\hat{b}\hat{d}} + \delta^{\hat{a}\hat{d}}n^{\hat{b}\hat{c}} - \delta^{\hat{b}\hat{d}} n^{\hat{a}\hat{c}}   \, .
\end{equation}
One can identify the $\mathfrak{so}(5)$ subalgebra by choosing one spatial direction, e.g. $\hat{b} = 6$, such that $n^{a 6} \equiv P^a$ are the momentum generators on S$^5$, and $n^{ab} \equiv J^{ab}$ are the angular momentum generators spanning $\mathfrak{so}(5)$, where $a, b, ... = 1, ..., 5$. 

Then the rescaling performed in \cite{Gomis:2005pg} is equivalent to the following one\footnote{We remark that in (\ref{our_resc1}), (\ref{our_resc2}) we have already taken into account the rescaling of the common AdS$_5$ and S$^5$ radius, which is imposed separately as an additional condition in \cite{Gomis:2005pg}.} 
\begin{eqnarray}
\label{our_resc1}
&&m^{\hat{i}\hat{j}} \rightarrow \omega \, m^{\hat{i}\hat{j}} \qquad\qquad \text{if} \quad \hat{i} \in \{ 0, 5, s_1 \} \quad \text{and} \quad \hat{j} \in \{s_2,s_3,s_4 \} \, , \\
\label{our_resc2}
&&m^{\hat{i}\hat{j}} \rightarrow  m^{\hat{i}\hat{j}}\hspace{1.9cm} \text{otherwise} \ ,  
\end{eqnarray}
where $s_1, s_2, s_3, s_4$ take a fixed value in the set $\{1, 2, 3, 4\}$, and
\begin{eqnarray}
\label{our_resc3}
n^{a 6} \rightarrow  \omega \, n^{a 6} \, , \qquad\qquad\qquad
n^{ab} \rightarrow n^{ab} \, . 
\end{eqnarray}
In the limit $\omega \rightarrow \infty$, the $\mathfrak{so}(4,2)$ algebra contracts to the Newton-Hooke algebra in 5-dimensions. Moreover, the subalgebra generated by $m^{\hat{i}\hat{j}}$ with $\hat{i}, \hat{j} \in \{ 0, 5, s_1 \} $ is the $\mathfrak{sl}(2, \mathbb{R})$ algebra, which will act as a global symmetry of the AdS$_2$ divergent part of the metric.

In the spinorial representation, the generators are
\begin{equation}
m^{ij} = \frac{1}{4} [\gamma^i, \gamma^j] \, , \qquad
m^{i5} = \frac{1}{2} \gamma^i \, , \qquad
n^{ab} =  \frac{1}{4} [\gamma^a, \gamma^b] \, , \qquad
n^{a 6} = \frac{i}{2} \gamma^a \, ,
\end{equation}
and we represent the gamma matrices as
\begin{eqnarray*}
%\nonumber
\gamma^1&=&{\scriptsize\left(
\begin{array}{cccc}
  0 & 0 & 0 & -1 \\
  0 & 0 & 1 & 0 \\
  0 & 1 & 0 & 0 \\
 -1 & 0 & 0 & 0
\end{array} \right)},\hspace{0.3in}
\gamma^2={\scriptsize\left(
\begin{array}{cccc}
  0 & 0 & 0 & i \\
  0 & 0 & i & 0 \\
  0 & -i & 0 & 0 \\
 -i & 0 & 0 & 0
\end{array} \right)},\hspace{0.3in}
\gamma^3={\scriptsize\left(
\begin{array}{cccc}
  0 & 0 & 1 & 0 \\
  0 & 0 & 0 & 1 \\
  1 & 0 & 0 & 0 \\
  0 & 1 & 0 & 0
\end{array} \right)}, \\
\nonumber \gamma^4&=&{\scriptsize\left(
\begin{array}{cccc}
  0 & 0 & -i & 0 \\
  0 & 0 & 0 & i \\
  i & 0  & 0 & 0\\
  0 & -i & 0 & 0
\end{array} \right)},\hspace{0.25in}~\gamma^5={\scriptsize \left(
\begin{array}{cccc}
  1 & 0 & 0 & 0 \\
  0 & 1 & 0 & 0 \\
  0 & 0 & -1 & 0 \\
   0 & 0 & 0 & -1
\end{array} \right)}\, ,\hspace{0.25in}
\gamma^0= i \gamma^5 \, .
\end{eqnarray*}

\subsection{Cartesian global coordinates}
Here we introduce the AdS$_5\times$S$^5$ analogue of the Cartesian coordinates for Minkowski space. 
The coset representative of $SO(4,2) \times SO(6)$ is of the type $g = $diag$(g_a, g_s)$, where $g_a$ and $g_s$ are coset representatives for $SO(4,2)$ and $SO(6)$ respectively. In the Cartesian global coordinates, $g_a$ and $g_s$ are taken as 
\begin{equation}
\label{cartesian_coset}
g_a = \Lambda(t) \cdot G(z) \ , \qquad\qquad\qquad
g_s = \Lambda(\phi) \cdot G(y) \ ,  
\end{equation}
where
\begin{eqnarray}
\notag
&&\hspace{1cm}\Lambda(t) = \exp ( t\, m^{05}) \ , \hspace{3.5cm}
\Lambda(\phi) = \exp ( \phi \, n^{56}) \ ,  \\
&&G(z) = \frac{1}{\sqrt{1- \frac{z^2}{4}}} \, ( \mathds{1} + z_i \,m^{i 5}) \ , \qquad\qquad
G(y) =  \frac{1}{\sqrt{1 + \frac{y^2}{4}}} \, ( \mathds{1} + y_i \, n^{i 6})\ ,
\end{eqnarray}
We remind the reader that we denote $z^2 \equiv z_i z^i$, where $z_i = z^i$, and $i = 1, ... , 4$. The same applies for the $y$ coordinates. Here $t$ is the global time in AdS$_5$ and $\phi$ is the angle describing the great circle in S$^5$.   
The AdS$_5\times$S$^5$ metric in these coordinates becomes
\begin{equation}
\dd s^2 = - \bigg(\frac{1+ \frac{z^2}{4}}{1-\frac{z^2}{4}}\bigg)^2 \dd t^2 + \frac{1}{(1-\frac{z^2}{4})^2} \dd z_i \dd z_i + \bigg(\frac{1 - \frac{y^2}{4}}{1 + \frac{y^2}{4}}\bigg)^2\dd \phi^2 + \frac{1}{(1+\frac{y^2}{4})^2} \dd y_i \dd y_i \ .
\end{equation}
In these coordinates, every choice of $s_1 \in \{1, 2, 3, 4\}$ entering in the rescaling (\ref{our_resc1}) and (\ref{our_resc2}) will give a sensible non-relativistic model based on String Newton-Cartan geometry.  For instance, one can take $s_1 = 1$, and the rescaling of coordinates that induce the algebra contraction (\ref{our_resc1}), (\ref{our_resc2}) and (\ref{our_resc3}) is 
\begin{equation}
t \rightarrow t \ ,  \qquad 
z_1 \rightarrow z_1 \ ,  \qquad
 z_m \rightarrow \frac{1}{\omega} z_m \ , \qquad
 \phi \rightarrow \frac{1}{\omega} \phi \ ,  \qquad
 y_i \rightarrow \frac{1}{\omega} y_i \ , 
\end{equation}
where $m = 2, 3,4$. 
By taking this choice of decomposition,  from the vielbeine expansion in powers of $\omega$ we read off the following String Newton-Cartan data 
\begin{eqnarray}
\notag
\tau_{\mu}{}^A &=& \text{diag}\bigg(-\frac{1+ (\frac{z_1}{2})^2}{1- (\frac{z_1}{2})^2},   \frac{1}{1 - (\frac{z_1}{2})^2}, 0, ... , 0 \bigg) \ , \\
m_{\mu}{}^A &=& \text{diag}\bigg(-\frac{z_m z_m}{2(1-(\frac{z_1}{2})^2)},  \frac{z_m z_m}{4(1-(\frac{z_1}{2})^2)}, 0, ... , 0 \bigg) \ , \\
\notag
e_{\mu}{}^a &=& \left(
\begin{array}{cccccccccc}
 0 & 0 & 0 & 0 & 0 & 0 & 0 & 0 & 0 & 0 \\
 0 & 0 & 0 & 0 & 0 & 0 & 0 & 0 & 0 & 0 \\
 0 & 0 & \frac{1}{1 - (\frac{z_1}{2})^2} & 0 & 0 & 0
   & 0 & 0 & 0 & 0 \\
 0 & 0 & 0 & \frac{1}{1 - (\frac{z_1}{2})^2} & 0 & 0
   & 0 & 0 & 0 & 0 \\
 0 & 0 & 0 & 0 & \frac{1}{1 - (\frac{z_1}{2})^2} & 0
   & 0 & 0 & 0 & 0 \\
 0 & 0 & 0 & 0 & 0 & 0 & 1 & 0 & 0 & 0 \\
 0 & 0 & 0 & 0 & 0 & 0 & 0 & 1 & 0 & 0 \\
 0 & 0 & 0 & 0 & 0 & 0 & 0 & 0 & 1 & 0 \\
 0 & 0 & 0 & 0 & 0 & 0 & 0 & 0 & 0 & 1 \\
 0 & 0 & 0 & 0 & 0 & 1 & 0 & 0 & 0 & 0 \\
\end{array}
\right) \ ,
\end{eqnarray}
where the ordering of the coordinates is $X^{\mu} =\{t,  z_1, z_2, z_3, z_4, \phi, y_1, y_2, y_3, y_4\}$. 

The \emph{longitudinal} metric $\tau_{\mu\nu}$ is
\begin{equation}
\tau_{\mu\nu} \dd X^{\mu} \dd X^{\nu} = - \bigg(\frac{1+ (\frac{z_1}{2})^2}{1- (\frac{z_1}{2})^2}\bigg)^2 \dd t^2 + \frac{1}{(1 - (\frac{z_1}{2})^2)^2} \dd z_1^2 \ ,
\end{equation}
and the \emph{boost invariant} tensor $H_{\mu\nu}$ is 
\begin{equation}
H_{\mu\nu} \dd X^{\mu} \dd X^{\nu} = - \frac{(1+ (\frac{z_1}{2})^2)z_m z_m}{(1- (\frac{z_1}{2})^2)^3} \, \dd t^2 
+ \frac{z_m z_m}{2(1- (\frac{z_1}{2})^2)^3}\,  \dd z_1^2 
+ \frac{1}{(1- (\frac{z_1}{2})^2)^2} \, \dd z_m \dd z_m + \dd \phi^2
+ \dd y_i \dd y_i \ .
\end{equation}
The closed B-field compensating the divergent part of the metric is 
\begin{equation}
B_{\mu\nu} \dd X^{\mu} \wedge \dd X^{\nu} = \omega^2 \frac{1+ (\frac{z_1}{2})^2}{(1 - (\frac{z_1}{2})^2)^2}  \dd t \wedge \dd z_1 \ . 
\end{equation}

\subsection{Polar global coordinates}
The set of polar global coordinates is obtained via the coset representative
\begin{equation}
g_a = \Lambda_a(t, \psi_1, \psi_2) \cdot \Theta_a(x) \cdot G_a(\rho) \ , \qquad\qquad
g_s = \Lambda_s(\phi,  \chi_1, \chi_2) \cdot \Theta_s(w) \cdot G_s(r) \ , 
\end{equation} 
where
\begin{eqnarray}
\notag
\Lambda_a(t, \psi_1, \psi_2)&=& \exp ( t \, m^{05} - \psi_1 \, m^{12} - \psi_2 \, m^{34} ) \ ,  \\
\notag
\Theta_a(x)&=& \exp ( \arcsin (x) \, m^{13} ) \ ,  \\
G_a(\rho) &=& \exp ( \arcsinh (\rho) \, m^{15} ) \ ,
\end{eqnarray}
and
\begin{eqnarray}
\notag
\Lambda_s(\phi,  \chi_1, \chi_2)&=& \exp ( \phi \, n^{56} - \chi_1 \, n^{12} - \chi_2 \, n^{34} ) \ ,  \\
\notag
\Theta_s(w)&=& \exp ( \arcsin (w) \, n^{13} ) \ ,  \\
G_s(r) &=& \exp ( \arcsin (r) \, n^{16} ) \ .
\end{eqnarray} 
In this set of coordinates the AdS$_5\times$S$^5$ metric reads as
\begin{equation}
\dd s^2 = \dd s^2_a + \dd s^2_s \ , 
\end{equation}  
where
\begin{eqnarray}
\notag
\dd s^2_a &=& -(1+ \rho^2) \dd t^2 + \frac{1}{1+ \rho^2} \dd \rho^2 +\frac{\rho^2}{1-x^2} \dd x^2+ \rho^2(1- x^2) \dd \psi_1^2 + \rho^2x^2  \dd \psi_2^2 \ , \\
\dd s^2_s &=& (1-r^2) \dd \phi^2 + \frac{1}{1-r^2} \dd r^2 + \frac{r^2}{1-w^2} \dd w^2 + r^2(1-w^2) \dd \chi_1^2 + r^2w^2 \dd \chi_2^2 \ .
\end{eqnarray}
In these coordinates, there is \emph{only one choice} of $s_1$ entering in (\ref{our_resc1}) and (\ref{our_resc2}) which produces a sensible non-relativistic action.  This choice is $s_1 = 1$, i.e. the index associated to the $\rho$ coordinate, and it produces an AdS$_2$ divergent part of the metric where the $\mathfrak{sl}(2, \mathbb{R})$ algebra coming from the contraction of $\mathfrak{so}(4,2)$ is acting on.  All other choices of $s_1$ will produce a divergent part of the metric which is the $\mathbb{R}$ metric, and this is not the non-relativistic model described in \cite{Gomis:2005pg}. 

Therefore, for the unique sensible choice $s_1 = 1$, the coordinate rescaling  induced by the algebra contraction (\ref{our_resc1}),  (\ref{our_resc2}) and (\ref{our_resc3}) is
\begin{eqnarray}
\notag
&&t \rightarrow t \ , \qquad
\rho \rightarrow \rho \, \qquad
\psi_2 \rightarrow \psi_2 \ , \qquad
\psi_1 \rightarrow \frac{1}{\omega} \psi_1 \ , \qquad
x \rightarrow  \frac{1}{\omega} x \ ,  \\
&&\chi_1 \rightarrow \chi_1 \ , \qquad
\chi_2 \rightarrow \chi_2 \ , \qquad
w \rightarrow w \ , \qquad
\phi \rightarrow \frac{1}{\omega} \phi \ , \qquad
r \rightarrow \frac{1}{\omega} r \ , \qquad
\end{eqnarray}
and from the vielbeine expansion we read off the String Newton-Cartan data
\begin{eqnarray}
\notag
\tau_{\mu}{}^A &=& \text{diag} ( - \sqrt{1+\rho^2},  \frac{1}{\sqrt{1+\rho^2}}, 0, ..., 0) \ ,  \\
m_{\mu}{}^A &=& 0 \ , \\
\notag
e_{\mu}{}^a &=& \left(
\begin{array}{cccccccccc}
 0 & 0 & 0 & 0 & 0 & 0 & 0 & 0 & 0 & 0 \\
 0 & 0 & 0 & 0 & 0 & 0 & 0 & 0 & 0 & 0 \\
 0 & 0 & 0 & \rho  & 0 & 0 & 0 & 0 & 0 & 0
   \\
 0 & 0 & -\rho  & 0 & 0 & 0 & 0 & 0 & 0 &
   0 \\
 0 & 0 & 0 & 0 & - \rho  x & 0 & 0 & 0 &
   0 & 0 \\
 0 & 0 & 0 & 0 & 0 & 0 & 1 & 0 & 0 & 0 \\
 0 & 0 & 0 & 0 & 0 & 0 & 0 & 0 & r
   \sqrt{1-w^2} & 0 \\
 0 & 0 & 0 & 0 & 0 & 0 & 0 &
   -\frac{r}{\sqrt{1-w^2}} & 0 & 0 \\
 0 & 0 & 0 & 0 & 0 & 0 & 0 & 0 & 0 & -r w
   \\
 0 & 0 & 0 & 0 & 0 & 1 & 0 & 0 & 0 & 0 \\
\end{array}
\right) \ ,
\end{eqnarray}
where the ordering of the coordinates is $X^{\mu} = \{t,  \rho, x, \psi_1, \psi_2, \phi, r, w, \chi_1, \chi_2\}$. 

The longitudinal metric $\tau_{\mu\nu}$ is 
\begin{equation}
\tau_{\mu\nu} \dd X^{\mu} \dd X^{\nu} = - (1+ \rho^2) \,  \dd t^2 + \frac{1}{1+ \rho^2} \, \dd \rho^2 \ ,
\end{equation}
and the boost invariant tensor $H_{\mu\nu}$ is 
\begin{equation}
\label{H_polar}
H_{\mu\nu} \dd X^{\mu} \dd X^{\nu} = \rho^2 ( \dd x^2 + \dd \psi_1^2 + x^2 \dd \psi_2^2) + \dd \phi^2 + \dd r^2 + r^2 \bigg( \frac{1}{1-w^2} \dd w^2 + (1-w^2) \dd \chi_1^2 + w^2 \dd \chi_2^2\bigg) \ .
\end{equation}
We remark that in this set of coordinates, the boost invariant tensor has not full rank, and therefore it may only be inverted (in the usual sense) only on a subspace of the coordinate system.

The closed B-field that compensates the divergent part of the metric is 
\begin{equation}
B_{\mu\nu} \dd X^{\mu} \wedge \dd X^{\nu} = \omega^2 \dd t \wedge \dd \rho \ . 
\end{equation}

\subsection{GGK coordinates}

The set of coordinates used by Gomis, Gomis and Kamimura \cite{Gomis:2005pg} is, in our convention,  given by the following choice of coset representative 
\begin{eqnarray}
\label{GGK_coset}
\notag
g_a &=& \exp ( x_1 \, m^{15} ) \exp ( x_0 \,m^{05}) \exp(x_a \,m^{a5}) \ , \\
g_s &=& \exp( \tilde{x}_{m'} \,n^{m'6} ) \ ,
\end{eqnarray}
where $a = 2, 3, 4$, and $m' = 1, ..., 5$. 
In this set of coordinates, the spacetime metric takes a complicated form given in \cite{Gomis:2005pg}, which we do not report here. 

Also in this set of coordinates, there is only one sensible choice of $s_1$ entering in (\ref{our_resc1}) and (\ref{our_resc2}). This is $s_1 = 1$, and intuitively it is because the choice of coset representative made here treats $x_1$ differently from the other AdS$_5$ coordinates.  The associated coordinates rescaling is 
\begin{equation}
x_0 \rightarrow x_0 \, \qquad
x_1 \rightarrow x_1\, \qquad
x_a \rightarrow \frac{1}{\omega} x_a \, \qquad
x_{m'} \rightarrow  \frac{1}{\omega} x_{m'} 
\end{equation}
The String Newton-Cartan data is 
\begin{eqnarray}
\notag
\tau_{\mu}{}^A &=& \text{diag} ( -1,  \cos x_0, 0, ..., 0) \ ,  \\
m_{\mu}{}^A &=& \frac{x_a x_a}{2} \text{diag} (-1,  \cos x_0, 0, ...., 0) \ , \\
\notag
e_{\mu}{}^a &=&\text{diag} (0,  0, 1, ...., 1) \ ,
\end{eqnarray}
where the ordering of the coordinates is $X^{\mu} = \{x_0, x_1, x_a, x_{m'}\}$. 

The longitudinal metric $\tau_{\mu\nu}$ is 
\begin{equation}
\tau_{\mu\nu} \dd X^{\mu} \dd X^{\nu} =  - \dd x_0^2 + \cos^2 x_0\, \dd x_1^2\ ,
\end{equation}
and the boost invariant tensor $H_{\mu\nu}$ is 
\begin{equation}
H_{\mu\nu} \dd X^{\mu} \dd X^{\nu} =  x_ax_a (-\dd x_0^2 + \cos^2x_0\, \dd x_1^2 )+ \dd x_a \dd x_a + \dd \tilde{x}_{m'} \dd \tilde{x}_{m'}\ .
\end{equation}
The closed B-field compensating the divergent part of the metric is 
\begin{equation}
B_{\mu\nu} \dd X^{\mu} \wedge \dd X^{\nu} = \omega^2 \cos x_0 \, \dd x_0 \wedge\dd x_1 \ . 
\end{equation}

\section{Cubic and Quartic Lagrangians}
\label{Appx_Cubic_Quartic}

Here we list the cubic and quartic Lagrangians entering in the expansion (\ref{S_expan}).  This shows that the perturbative expansion of the action in the large string tension parameter is well-defined.  They all show a manifest $SO(3)\times SO(4)$ symmetry.

\begin{eqnarray}
\notag
\mathcal{L}_3 &=& -\frac{s}{4}\bigg[2 p_{\tz} (p_{\tz}
   \tilde{z}' +p_{z_m} z_m' + p_{y_i} y_i'
   )-\tilde{z}'
   (p_{\tz}^2+p_{z_m}^2+p_{y_i}^2+\tilde{z}'^2 + z_m'^2 + y_i'^2)\bigg] \ , \\
\end{eqnarray}

\begin{eqnarray}
\notag
\hat{\mathcal{L}}_3 &=& -\frac{s}{16}\bigg[ (\sigma-\pi)
   ^2 \tilde{z}' ( p_{\tz}^2+3 p_{z_m}^2 + p_{y_i}^2) +4\tilde{z}' \tilde{z}^2+18 \tilde{z}'z_m^2 +(\sigma-\pi)^2\tilde{z}' (-z_m'^2+y_i'^2)\\
   \notag
   &+&2(\sigma-\pi)^2 p_{\tz}p_{z_m}z_m'
  +2(\sigma-\pi)^2 p_{\tz}p_{y_i}y_i'+3(\sigma-\pi) ^2 \tilde{z}'^3\bigg] \\
 &-&\frac{(\sigma-\pi)}{8} \tilde{z} \left( 3 p_{\tz}^2-3 p_{z_m}^2+ p_{y_i}^2 + 5 z_m'^2 + y_i'^2 - \tilde{z}'^2 \right) \ ,
\end{eqnarray}

\begin{eqnarray}
\notag
\mathcal{L}_4 &=& -\frac{1}{32}\bigg[\tilde{z}'^2 (-2 p_{\tz}^2+6p_{z_m}^2 + 6p_{y_i}^2+6z_m'^2 + 6y_i'^2) +(p_{\tz}^2 + p_{z_m}^2 + p_{y_i}^2+z_m'^2 + y_i'^2)^2\\
   &-&8 p_{\tz}\tilde{z}' (p_{z_m}z_m'+p_{y_i}y_i')+5\tilde{z}'^4 \bigg] \ , 
\end{eqnarray}

\begin{eqnarray}
\notag
\hat{\mathcal{L}}_4 &=&-\frac{1}{32}\bigg[ \tz^2( 6 p_{\tz}^2  - 6 p_{z_m}^2 + 2 p_{y_i}^2) + 3 z_{m}^2 (-p_{\tz}^2 + p_{z_m}^2 + p_{y_i}^2) + 2 (\sigma-\pi) s \tz \tz'^3 \\
\notag
&+& (\sigma-\pi)^2 (p_{\tz}^2 - p_{z_m}^2)(p_{\tz}^2 + p_{z_m}^2 + p_{y_i}^2) + (\sigma-\pi)^2 ( z_m'^4 - 5 \tz'^4) \\
\notag
&+& 12 (\sigma-\pi) s \tz p_{\tz} (p_{z_m} z_m' + p_{y_i} y_i') 
+ y_i'^2 \left(2\tz^2 + 3 z_m^2 + (\sigma-\pi)^2(p_{\tz}^2 - p_{z_m}^2) \right) \\
\notag
&+& 2 (\sigma-\pi) s \tz \tz' \left( 3p_{\tz}^2 + 3p_{z_m}^2 - p_{y_i}^2 - 5 z_m'^2 - y_i'^2\right) \\
\notag
&+& z_m^2 \left(10 \tz^2 + 3 z_m^2 + (\sigma-\pi)^2(2p_{\tz}^2 + p_{y_i}^2 + y_i'^2) \right) \\
&-& \tz^2 \left( 2\tz^2 - 9 z_m^2 + 3 (\sigma-\pi)^2 (2p_{z_m}^2 + p_{y_i}^2 + y_i'^2)\right) \bigg] \ . 
\end{eqnarray}

\end{appendices}

%%%%%%%%%%%%%%%% BIBLIOGRAPHY %%%%%%%%%%%%%%%%

\bibliographystyle{nb}

\bibliography{Biblio.bib}

\end{document}